\documentclass[aps, prc, onecolumn, 
showpacs, showkeys, 
print, preprintnumbers, 
groupedaddress,
nofootinbib
]{revtex4-2}

\usepackage[utf8]{inputenc}
\usepackage[T1]{fontenc}
\usepackage{lmodern}
\usepackage{microtype}
\usepackage[english]{babel}

\usepackage{amssymb,amsbsy,amsmath,amsfonts,amsthm}
\usepackage{yhmath} 

\usepackage{slashed}
\usepackage{bm}
\usepackage{mathptmx}  

\usepackage[dvipsnames,table,xcdraw]{xcolor}
\definecolor{lightyellow}{RGB}{255,250,205}

\usepackage{graphicx}
\usepackage{epstopdf}

\usepackage{array}        

\usepackage{multirow}
\usepackage{longtable}

\usepackage{url}
\usepackage{orcidlink}

\usepackage{comment}
\usepackage{diagbox}
\usepackage{soul}

\usepackage{hyperref} 
\hypersetup{ 
  colorlinks=true,
  linkcolor=blue,  
  citecolor=cyan,
  urlcolor=red,           
}

\begin{document}

\title{
Nucleon-nucleon scattering up to next-to-leading order in manifestly Lorentz-invariant chiral effective field theory: low phases and the deuteron}

\author{Xiu-Lei Ren\orcidlink{0000-0002-5138-7415}}
\email[]{xiulei.ren@sdu.edu.cn}
\affiliation{Shandong Provincial Key Laboratory of Nuclear Science, Nuclear Energy Technology and Comprehensive Utilization, School of Nuclear Science, Energy and Power Engineering, Shandong University, Jinan 250061, China}

\author{E.~Epelbaum\orcidlink{0000-0002-7613-0210}
}
 \affiliation{Institut f\"ur Theoretische Physik II, Ruhr-Universit\"at Bochum,  D-44780 Bochum,
 Germany}
\author{J.~Gegelia\orcidlink{0000-0002-5720-9978}
}
 \affiliation{Institut f\"ur Theoretische Physik II, Ruhr-Universit\"at Bochum,  D-44780 Bochum,
 Germany}
\affiliation{Tbilisi State  University,  0186 Tbilisi, Georgia}

\begin{abstract}
Recently the nucleon-nucleon interaction derived using time-ordered perturbation theory in manifestly Lorentz-invariant chiral effective field theory was shown to yield promising results for peripheral neutron-proton scattering. In this work we study low partial waves at next-to-leading order by treating the potential non-perturbatively in the scattering equation.  Reasonable description of the phase shifts in the $S$ and $P$ waves as well as the deuteron properties is observed, which can be regarded as a feasibility study for the application of our formalism to the few- and many-body calculations.      
\end{abstract}

\date{\today}
\maketitle

\section{Introduction}

Nucleon-nucleon and few-nucleon scattering observables have been extensively studied in the chiral effective field theory (ChEFT) since Weinberg's seminal works in 1990s~\cite{Weinberg:1990rz,Weinberg:1991um}. Nowadays, this remains an active research field with ongoing efforts focused on the derivation of high precision chiral potentials including three-/four-body forces, understanding renormalization issues in the non-perturbative regime and  providing reliable input for {\it ab initio} studies of finite nuclei and dense nuclear matter. 
In particular, the nucleon-nucleon interaction has been formulated up to fifth order of the chiral expansion within the non-relativistic (heavy-baryon) scheme of ChEFT~\cite{Entem:2017gor,Reinert:2017usi}. The development path and some recent progress have been covered in the reviews~\cite{Bedaque:2002mn,Epelbaum:2005pn,Epelbaum:2008ga,Machleidt:2011zz,Hammer:2019poc,Epelbaum:2019kcf,LE.Marcucci:2021,A.Kievsky:2022,Tews:2022yfb} and references therein. 

Besides the canonical method for formulating nuclear forces within the non-relativistic ChEFT, it is worth  pointing out that relativistic corrections can alter the ultraviolet (UV) behavior of the effective potential and the scattering equation, while maintaining the same infrared behavior as in the non-relativistic formalism. 

To exploit different UV behavior due to relativistic corrections, a modified Weinberg approach was proposed in Ref.~\cite{Epelbaum:2012ua} by using the manifestly Lorentz-invariant effective Lagrangian. Through iterating the standard leading order (LO) effective potential in the Kadyshevsky equation~\cite{Kadyshevsky:1967rs} (instead of the Lippmann-Schwinger equation), it results in a perturbatively renormalizable framework for NN scattering, requiring only an additional contact term in the $^3P_0$ channel at LO.

Inspired by the modified Weinberg approach, time-ordered perturbation theory (TOPT) has been applied in Ref.~\cite{Baru:2019ndr} to systematically formulate the nuclear force using the covariant chiral Lagrangians~\cite{Ren:2019qow,Ren:2022glg}~\footnote{There are different strategies to study the NN interaction up to NNLO within the relativistic ChEFT~\cite{Higa:2003jk,Higa:2003sz,Ren:2016jna,Lu:2021gsb}.}. The effective potential, defined as a sum of two-nucleon-irreducible time-ordered diagrams, can be improved order by order. 
The corresponding three-dimensional scattering equation is obtained self-consistently within the TOPT framework. The resulting equation coincides in form with the Kadyshevsky equation~\cite{Kadyshevsky:1967rs}  and the Thompson-I equation~\cite{Thompson:1970wt}. We believe that this formulation, firmly rooted in TOPT, provides a transparent framework for investigating the renormalization issue. Note that the formulation we employ here follows the lines of the heavy-baryon scheme within the TOPT framework of Refs.~\cite{Weinberg:1990rz,Ordonez:1995rz}. 

The LO study of the NN scattering provides a renormalizable framework~\cite{Baru:2019ndr}, where all partial waves have unique solutions given by the scattering equation. The same argument has been extended to hyperon-nucleon scattering in Ref.~\cite{Ren:2019qow}. Encouraged by this fact, in Ref.~\cite{Ren:2022glg} we carried out the calculation of the two-pion exchange (TPE) potentials up to next-to-next-to-leading order (NNLO). A reasonable description of peripheral partial waves is achieved by employing the Born series up to one-loop level. The improvement for $D$ waves is observed in comparison with the non-relativistic counterparts. 

In this work, we study low partial waves, i.e. the $S$ and $P$ waves, in the non-perturbative regime of NN scattering up to next-to-leading order (NLO). As a first step, we obtain the NN scattering amplitude by iterating the full NLO potential in the scattering equation. The low-energy constants (LECs) are determined  from fitting to the empirical phase shifts of $S$ and $P$ waves. The prediction for $D$ waves and higher partial waves is presented to  quantify non-perturbative effects. We also give the predictions of deuteron properties up to NLO. 

The paper is organized as follows: in Sec.~\ref{sec2}, we briefly introduce the chiral effective potential up to NLO, present the scattering equation and define the corresponding phase shift.  Next, we fix the LECs by describing the phase shifts of the $S$ and $P$ waves, and present the predictions for the higher partial-wave phase shifts and the deuteron properties in Sec.~\ref{sec3}. The results of our work are summarized in Sec.~\ref{sec4}.

\section{Formalism}\label{sec2}
\subsection{Chiral potential up to NLO}
The covariant effective Lagrangian of the NN sector at lowest order reads as 
\begin{equation}
   \mathcal{L}_{NN}^{(0)}= -\frac{1}{2} \biggl[ C_S(\bar{\Psi} \Psi)^2 + C_A(\bar{\Psi} \gamma_5 \Psi)^2 + C_V(\bar{\Psi} \gamma_\mu \Psi)^2 
   + C_{AV}(\bar{\Psi} \gamma_\mu\gamma_5 \Psi)^2 + C_T(\bar{\Psi} \sigma_{\mu\nu} \Psi)^2 \biggr],	
\end{equation}
where $C_{S,A,V,AV,T}$ are low-energy constants. The corresponding contact interaction potential is 
\begin{equation}\label{Eq:LOcontact}
\begin{aligned}
    V_{\mathrm C}^{(0)}(\bm{p}',\bm{p}) &   = C_S\, \bar{u}(\bm{p}') u(\bm{p}) \,  \bar{u}(-\bm{p}') u(-\bm{p})  + C_A\, \bar{u}(\bm{p}')\gamma_5 u(\bm{p})  \, \bar{u}(-\bm{p}')\gamma_5 u(-\bm{p}) + C_V\, \bar{u}(\bm{p}')\gamma_\mu u(\bm{p}) \, \bar{u}(-\bm{p}')\gamma^\mu u(-\bm{p}) \\
    &\quad  + C_{AV}\, \bar{u}(\bm{p}')\gamma_\mu\gamma_5 u(\bm{p}) \, \bar{u}(-\bm{p}')\gamma^\mu\gamma_5 u(-\bm{p}) 
    + C_T\, \bar{u}(\bm{p}')\sigma_{\mu\nu} u(\bm{p}) \, \bar{u}(-\bm{p}')\sigma^{\mu\nu} u(-\bm{p}) ,
\end{aligned}	
\end{equation}
where the spin indices of the Dirac spinors are suppressed, and $\bm{p}~(\bm{p'})$ is the initial (final) relative three-momentum in the center-of-momentum system. 
Since we apply the standard Weinberg power counting to organize the NN potential up to NLO, the expansion of the nucleon energies should be up to $\mathcal{O}(p^4)$ at least, 
\begin{equation}\label{Eq:DiracExpand}
\sqrt{\omega_p+m_N} = \sqrt{2m_N} + \frac{p^2}{4\sqrt{2}\, m_N^{3/2}} + \mathcal{O}(p^4),	
\end{equation}
where $\omega_p$ is the energy of nucleon with the magnitude of momentum $p=|\bm{p}|$, $\omega_p:=\sqrt{p^2+m_N^2}$.  
In agreement to the conventional Weinberg power counting, the LO contribution of the contact interaction terms of the effective Lagrangian to the potential $V_C^{(0)}$, i.e., not suppressed by factors of $1/m_N$, is reduced to $(C_S+C_V)-(C_{AV}-2C_T)\bm{\sigma}_1\cdot \bm{\sigma}_2$. It is obtained by taking the first term $\sqrt{2m_N}$ on the right-hand side of Eq.~\eqref{Eq:DiracExpand}. This result is consistent with the non-relativistic contact term at LO. 
 The potential $V_C^{(0)}$ also contains the NLO corrections, i.e. the momentum-dependent part as a function of $\bm{p}$ and $\bm{p}'$. This contribution can be obtained by keeping the second term in the righthand side of Eq.~\eqref{Eq:DiracExpand}.
For simplicity, we include all higher orders for contact potential in Eq.~\eqref{Eq:LOcontact} by keeping the full form of Dirac spinors.

The NN covariant Lagrangians at NLO with two derivatives acting on the nucleon fields have been studied in Refs.~\cite{Girlanda:2010ya,Xiao:2018jot,Filandri:2023qio}. 
Although the minimal set of operators at $\mathcal{O}(p^2)$ is not finally worked out, the non-relativistic reduction of those covariant Lagrangians is consistent with the conventional NN Lagrangians in the heavy-baryon scheme. Since we are working with the Weinberg power counting, the NLO correction of the contact interaction terms from the second order NN covariant Lagrangian is the same as its non-relativistic counterpart after expanding the nucleon energies as $\sqrt{\omega_p+m_N} = \sqrt{2m_N} + \mathcal{O}(p^2)$. Thus, the NLO contact interaction potential is written as 
\begin{equation}\label{Eq:NLOcontact}
\begin{aligned}
V_\mathrm{C}^{(2)} &= C_{1} \bm{q}^{2}+C_{2} \bm{P}^{2} +\left(C_{3} \bm{q}^{2}+C_{4} \bm{P}^{2}\right)\left(\bm{\sigma}_{1} \cdot \bm{\sigma}_{2}\right)+\frac{i}{2}C_{5} \left(\bm{\sigma}_{1}+\bm{\sigma}_{2}\right) \cdot \bm{n} \nonumber\\
&\quad +C_{6}\left(\bm{q} \cdot \bm{\sigma}_{1}\right)\left(\bm{q} \cdot \bm{\sigma}_{2}\right)+C_{7}\left(\bm{P} \cdot \bm{\sigma}_{1}\right)\left(\bm{P} \cdot \bm{\sigma}_{2}\right) ,
\end{aligned}
\end{equation}
with $\bm{q}=\bm{p}'-\bm{p}$, $\bm{P} = (\bm{p}+\bm{p}')/2$,  and $\bm{n}=\bm{p}\times \bm{p}'=\bm{P}\times \bm{q}$. The LECs $C_{1,...,7}$ are unknown and can be fixed by the NN scattering phase shifts or observables. 

Next, we follow the standard procedure to perform the partial wave projection of the contact terms 
\begin{equation}
  V_\mathrm{C} = V_\mathrm{C}^{(0)} + V_\mathrm{C}^{(2)},
\end{equation}
in the $|lsj\rangle$ basis, where $l$ denotes the total orbital angular momentum, $s$ the total spin, and $j$ the total angular momentum. Finally we have the contributions of contact terms to the low partial waves,
\begin{equation}\label{Eq:contact}
\begin{aligned}
	V(^1S_0)&= \xi_{N} \biggl[ \tilde{C}_{^1 S_0} + \tilde{C}_{^1 S_0}R_{p}^{2} R_{p^{\prime}}^{2}  
+ C_{^1S_0}  \left( R_p^2 +R_{p'}^2 \right) \biggr], \\
V(^3S_1) 
 &= \xi_{N} \left[ \tilde{C}_{^3S_1}  + \frac{\tilde{C}_{^3S_1}}{9} R_{p}^{2} R_{p^{\prime}}^{2} 
 +  \frac{C_{^3S_1}}{9} \left( R_p^2 + R_{p'}^2 \right) \right], \\
 V(^3S_1-^3D_1)  &=  \xi_{N}  \left[ C_{\epsilon_1}\, R_{p}^2 + \frac{2 \sqrt{2}}{9}\, \tilde{C}_{^3S_1}  R_{p}^{2} R_{p^{\prime}}^{2} \right],\\
  V(^3D_1-^3S_1) &=  \xi_{N} \left[ C_{\epsilon_1} \, R_{p^{\prime}}^{2} + \frac{2 \sqrt{2} }{9} \tilde{C}_{^3S_1}
   R_{p}^{2} R_{p^{\prime}}^{2} \right],\\
   V(^3D_1) &= \frac{8 \xi_{N}}{9} \tilde{C}_{^3S_1} R_{p}^{2} R_{p^{\prime}}^{2}, \\
V(^3P_0) &=  C_{^3P_0}\, p\, p' , \\
V(^1P_1) &=  C_{^1P_1} \, p\, p',\\
V(^3P_1) &=  C_{^3P_1} \, p\, p',\\
 V(^3P_2)  &=C_{^3P_2} \,  p\, p',
\end{aligned}
\end{equation}
where 
\begin{equation}
\xi_{N}
= \frac{(\omega_p+m_N)(\omega_{p'}+m_N)}{4m_N^2}, \quad R_{p}=\frac{p}{\omega_p+m_N}, \text { and } R_{p^{\prime}}=\frac{p^{\prime}}{\omega_{p'}+m_N}.
\end{equation}
It is worth to note that we have 9 parameters (as the combinations of LECs $C_{S,A,V,AV,T}$ and $C_{1,...,7}$) to be fixed, 
\begin{equation}
\begin{aligned}
\tilde{C}_{^1S_0} &= 4 \pi \left(C_{S}+C_{V}+3 C_{A V}-6 C_{T}\right),\\
\tilde{C}_{^3S_1} &= 4 \pi \left(C_{S}+C_{V}-C_{A V}+2 C_{T}\right),\\
	C_{^1S_0} &= 4 \pi \left(3 C_{V}+C_{A}+C_{A V}-6 C_{T}\right) +  4m_N^2 \pi \, ( 4C_1 + C_2 -12C_3
-3C_4 -4C_6 -C_7) , \\
	C_{^3S_1} &= 12 \pi \left(C_{V}-C_{A}-C_{A V}-2 C_{T}\right) + 12 m_N^2 \pi \, ( 12C_1 + 3C_2 +12C_3
+3C_4 +4C_6 +C_7) , \\
C_{\varepsilon_1} &=  \frac{8\sqrt{2}}{3}\pi  \left(C_{V}-C_{A}-C_{A V}-2 C_{T}\right) + 4m_N^2 \frac{2\sqrt{2}\pi}{3} \, ( 4C_6 + C_7)   ,\\
C_{^3P_0} &= -\frac{2\pi}{m_N^2} \left(C_{S}-4 C_{V}+C_{A}-4 C_{A V}-12C_T\right)  + \frac{2\pi}{3} \, ( -4C_1 + C_2 - 4C_3
+C_4 + 4C_5 +12C_6 - 3C_7)   ,\\
C_{^1P_1} &= -\frac{2\pi}{3m_N^2} \left(C_{S}+C_{A}+4C_T\right)  +  \frac{2\pi}{3} \, ( -4C_1 + C_2 +12C_3
-3C_4 +4C_6 -C_7) ,\\
C_{^3P_1} &= -\frac{4\pi}{3m_N^2} \left(C_{S}-2 C_{V}-C_{A}+2 C_{A V}\right) +  \frac{2\pi}{3} \, ( -4C_1 + C_2 - 4C_3
+C_4 + 2C_5 +4C_6 + C_7) ,\\
C_{^3P_2} &= \frac{2\pi}{3} \, ( -4C_1 + C_2 - 4C_3
+C_4 + 2C_5 ),
\end{aligned}
\end{equation}
which is the same number of contact terms as in the non-relativistic case up to NLO. After performing the non-relativistic expansion of the $^1S_0$ and the coupled $^3S_1$-$^3D_1$ channels, one finds the following relations between the LECs in our scheme and in the heavy-baryon approach: 
 \begin{equation}
 	 \tilde{C}_{^1S_0}^{NR} = \tilde{C}_{^1 S_0},  \quad  C_{^1S_0}^{NR} = \frac{(\tilde{C}_{^1S_0}+C_{^1S_0})}{4m_N^2}, 
 	 \quad 
 	 \tilde{C}_{^3S_1}^{NR} = \tilde{C}_{^3 S_1}, \quad  C_{^3S_1}^{NR} = \frac{(9\tilde{C}_{^3S_1}+C_{^3S_1})}{36m_N^2},\quad 
 	 C_{\epsilon_1}^{NR} =\frac{ C_{\epsilon_1} }{4m_N^2}.
 \end{equation}

The pion-exchange potential up to NLO has been explicitly presented in Ref.~\cite{Ren:2022glg}. For completeness, we briefly summarize the pertinent formalism in the following. The chiral pion-exchange potential is written as  
\begin{equation}
V_{\pi} = V_{1\pi}^{(0)} + V_{2\pi}^{(2)} ,
\end{equation}
where the one-pion exchange (OPE) potential
 \begin{equation}
   V^{(0)}_{1\pi}(M_\pi) = -\frac{g_A^2}{4 f_\pi^2} \frac{1}{\bm{q}^2+M_\pi^2}\left(\bar{u}_3 \gamma_\mu \gamma_5 q^\mu u_1\right)\left(\bar{u}_4 \gamma_\nu \gamma_5 q^\nu u_2\right)
\end{equation}
is given in Ref.~\cite{Ren:2022glg}.
The TPE potentials $V_{2\pi}^{(2)}$ up to NLO have been worked out in Ref.~\cite{Ren:2022glg}, the corresponding expressions are presented in Eqs.~(13), (14), (17), (20), and (22)~of that work.

\subsection{Scattering equation and phase shift}
To obtain the scattering $T$-matrix for the $S$ and $P$ partial waves, we need to treat the NLO potential, $V=V^{C}+V^{\pi}$, non-perturbatively by solving the scattering equation in the partial wave basis,
\begin{equation}\label{Eq:kady}
\begin{aligned}
T_{ll'}^{sj}\left(p^{\prime}, p\right)&=V_{ll'}^{sj}\left(p^{\prime}, p\right) \\
&+\sum\limits_{l''}\int \frac{{p''}^2d p''}{(2 \pi)^{3}} V_{ll''}^{sj}\left(p^{\prime}, p''\right) \frac{m_{N}^{2}}{2 ({p''}^2+m_N^2)} \frac{1}{\sqrt{p^2+m_N^2}-\sqrt{{p''}^2+m_N^2}+i \varepsilon} T_{l''l'}^{sj}(p'', p).
\end{aligned}	
\end{equation}
The partial wave $S$ matrix is related to the on-shell $T$ matrix by \begin{equation}
  S_{l'l}^{sj}(p) = \delta_{l'l} - \frac{i}{8\pi^2} \frac{ p\, m_N^2}{\omega_p} T_{l'l}^{sj}(p).
\end{equation}
The phase shifts can be obtained via the parameterization of $S$ matrix as:
\begin{itemize}
\item Single channel case 
\begin{equation} 
  S_{j j}^{0 j}=\exp(2 i \delta_j^{0 j}), \quad S_{j j}^{1 j}=\exp(2 i \delta_j^{1 j}),
\end{equation}
with the notation of phase shift $\delta_l^{sj}$.
\item Coupled channel case ($j>0$)\\
We use the so-called Stapp (bar-) parameterization~\cite{Stapp:1956mz},
\begin{equation}
\begin{aligned}
S & =\left(\begin{array}{cc}
S_{j-1 j-1}^{1 j} & S_{j-1 j+1}^{1 j} \\
S_{j+1 j-1}^{1 j} & S_{j+1 j+1}^{1 j}
\end{array}\right) \\
& =\left(\begin{array}{cc}
\cos (2 \epsilon_j) \exp \left(2 i \delta_{j-1}^{1 j}\right) & i \sin (2 \epsilon_j) \exp \left(i \delta_{j-1}^{1 j}+i \delta_{j+1}^{1 j}\right) \\
i \sin (2 \epsilon_j) \exp \left(i \delta_{j-1}^{1 j}+i \delta_{j+1}^{1 j}\right) & \cos (2 \epsilon_j) \exp \left(2 i \delta_{j+1}^{1 j}\right)
\end{array}\right) ,
\end{aligned}
\end{equation}
with the mixing parameter $\epsilon_j$. 
\end{itemize}

In solving the scattering equation, one usually introduces a regulator function to the chiral potential in order to avoid the UV divergences when the momentum of integration goes to infinity. Although our LO study can avoid such cutoff artefacts, the TPE corrections are well known to be more singular than the OPE potential. Thus, we need to introduce a finite momentum cutoff $\Lambda$ to iterate the TPE potentials in the scattering equation. For the current study, we utilize a simple Gaussian momentum-space regulator~\cite{Epelbaum:1999dj} for both the contact terms and the pion-exchange potentials, 
\begin{equation}\label{Eq:GaussReg}
   V(p', p) \to 	e^{-{p'}^{2n}/\Lambda^{2n}}V(p', p) e^{-{p}^{2n}/\Lambda^{2n}},
\end{equation}
with $n=2$ to make sure the introduced error is beyond the accuracy of NLO.

\section{Results and discussion}\label{sec3}
In this section, we present the description of partial-wave phase shifts and the deuteron properties up to NLO. The parameters used in the subsequent calculations are: the pion masses $M_{\pi^\pm}=139.570$ MeV, $M_{\pi^0}=134.977$ MeV, and the average pion mass appearing in the TPE potentials $M_\pi=138.0$ MeV; the nucleon mass $m_N = 938.918$ MeV; the pion decay constant $f_\pi = 92.4$ MeV; the axial coupling $g_A$ fixed as $1.29$ at NLO  to take into account the Goldberger-Treiman discrepancy~\cite{Goldberger:1958tr}.  

In order to determine the contact terms, i.e. the values of LECs $C$ and $\tilde{C}$ defined in Eq.~\eqref{Eq:contact}, 
we employ a simple method to reproduce the phase shifts of low partial waves given by Nijmegen group. 
The corresponding chi-square is defined as 
\begin{equation}
  \chi^2 = \sum\limits_i \frac{(\delta_i^\mathrm{Th}-\delta_i^\mathrm{Nij})^2}{{\Delta\delta_i^\mathrm{Nij}}^2},
\end{equation}
where the sum index $i$ denotes the different laboratory energies $E_\mathrm{lab}=1$,  $5$, $10$, $25$, $50$, $100$, $150$, $200$ MeV, as employed in the Nijmegen partial wave analysis (PWA)~\cite{Stoks:1993tb}. 
As to the uncertainties of phase shifts, the Nijmegen results only have statistical errors and do not include systematic uncertainties. To have a meaningful fit, we employ a more reasonable estimation of errors of phase shifts $\Delta \delta_i^\mathrm{Nij}$ given in Ref.~\cite{Epelbaum:2014efa}, where the systematic uncertainties are included by considering the variations of Nijmegen I, II, and Reid 93 NN potentials~\cite{Stoks:1994wp}. 
We perform fits of phase shifts up to $E_\mathrm{lab}\leq 50$ MeV at NLO. 
The corresponding values of LECs $C$ and $\tilde{C}$ are listed in Table~\ref{Tab:LECs}, 
where the momentum cutoff $\Lambda$ in Eq.~\eqref{Eq:GaussReg} is varied from $400$ MeV to $650$ MeV. 

\begin{table}[t]
\caption{Values of LECs $\tilde{C}$ and $C$ at NLO varying $\Lambda=400$ -- $650$ MeV. The values of $\tilde{C}$ are in $10^{4}$ GeV$^{-2}$, $C_{3P0,1P1,3P1,3P2}$ are in $10^{4}$ GeV$^{-4}$, and $C_{\epsilon_1}$, $C_{1S0,3S1}$ are in units of $10^{6}$ GeV$^{-2}$.} 
\label{Tab:LECs}
\begin{ruledtabular}
\begin{tabular}{c|rrrrrr}
\diagbox{LECs}{$\Lambda$ (MeV)} &  $400$      & $450$       & $500$        & $550$      & $600$      & $650$   \\
\hline  
 $\tilde{C}_{^1S_0}$  & $-0.182$   & $-0.182$   & $-0.179$   & $-0.171$  & $-0.144$  & $-0.00534$ \\
 $C_{1 S 0}$            & $0.0153$ & $0.0168$  & $0.0186$  & $0.0215$ & $0.0269$ & $0.0408$  \\
 \hline 
 $\tilde{C}_{^3S_1}$  & $-0.208$  & $-0.199 $  & $-0.191$  & $-0.175$  & $-0.124$   & $0.162$ \\
 $C_{^3S_1}$            &  $0.514$  & $0.518 $   & $ 0.531$  & $0.567 $  & $0.654$    & $0.900$ \\
 $C_{\epsilon_1}$   & $-0.0266$  & $-0.0261$ & $-0.0260$ & $-0.0267$ & $-0.0289$  & $-0.0351$ \\
\hline
$C_{^3P_0}$ & $0.357$ & $0.377$ & $0.405$ & $0.450$ & $0.522$ & $0.646$ \\
\hline
$C_{^1P_1}$ & $0.230$ & $0.219$ & $0.220$ & $0.229$ & $0.245$ & $0.268$ \\
\hline
$C_{^3P_1}$ & $0.0317$ & $0.0118$ & $0.00390$ & $0.00211$ & $0.00373$ & $0.00750$ \\
\hline
$C_{^3P_2}$ & $-0.406$ & $-0.370$ & $-0.338$ & $-0.308$ & $-0.281$ &$ -0.256$ \\
\end{tabular}
\end{ruledtabular}
\end{table}

\begin{figure}[t]
  \includegraphics[width=0.8\textwidth]{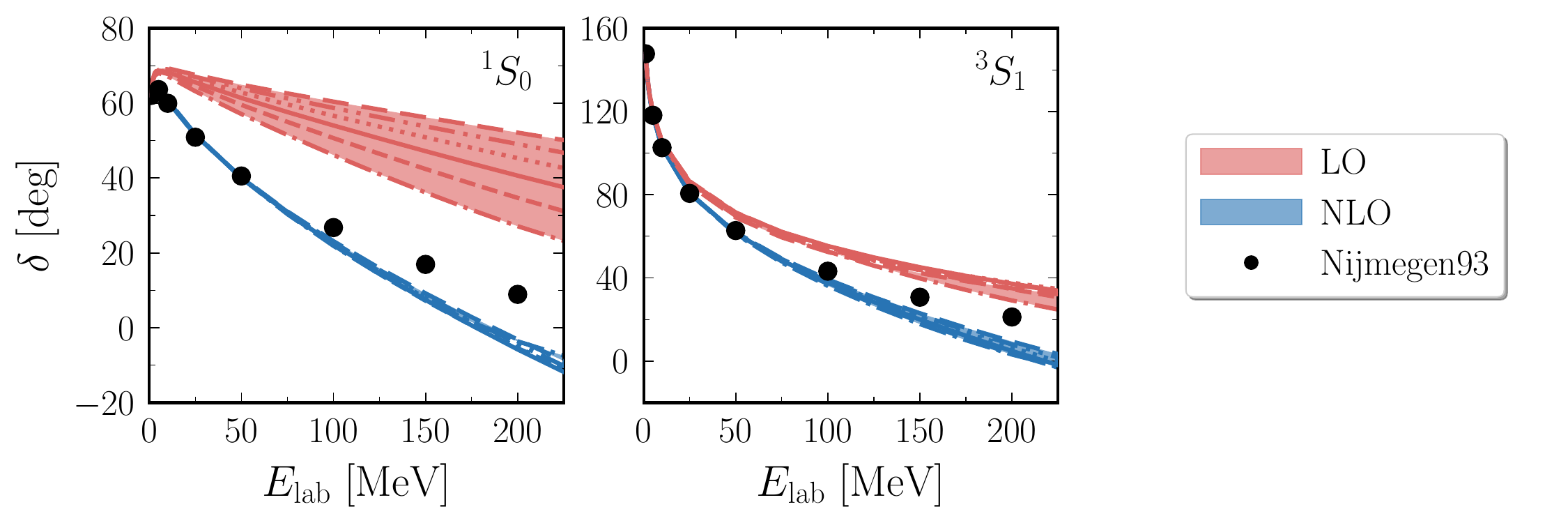}
  \caption{$S$-wave neutron-proton phase shifts versus the laboratory energy. The light-red and blue bands correspond to the LO and NLO results. The cuff-off $\Lambda$ in the scattering equation is varied in the range $\Lambda=400$ -- $650$ MeV. The filled circles represent the results of the Nijmegen
  PWA~\cite{Stoks:1993tb}. 
  }
  \label{Fig:Swave}
\end{figure}

\subsection{S-waves}
Having  determined the two LECs of the $^1S_0$ wave ($\tilde{C}_{^1S_0}$ and  $C_{^1S_0}$) by fitting the phase shifts at NLO, we present the predictions of $^1S_0$ phases for $E_\mathrm{lab}$ up to $200$ MeV in Fig.~\ref{Fig:Swave}. The LO result is also presented for comparison, where the single parameter is determined by the phase shift with $E_\mathrm{lab}=1$ MeV.
A clear improvement is found from LO to NLO. Similar behavior is also observed in the $^3S_1$ partial wave, where the LECs are determined by fitting the $^3S_1$, $^3D_1$ phases and the mixing parameter $\epsilon_1$. 
The light-red and blue bands denote the dependence on the momentum cutoff ($\Lambda=400$ -- $650$ MeV) applied in solving the scattering equation. The width of blue bands is small at NLO, which indicates the relatively small dependence on the momentum cutoff $\Lambda$.

Besides, one can predict the scattering length $a$ and effective range $r$, which are given in terms of the effective range expansion of the $S$-wave  
\begin{equation}
  p\cot(\delta) = - \frac{1}{a} + \frac{1}{2}r\,p^2 +  \cdots,	
\end{equation}
where $p$ is the nucleon center-of-mass momentum. In Table~\ref{Tab:1S0-3S1ERE}, we present the effective range coefficients up to NLO in the $^1S_0$ and $^3S_1$ channels. For comparison, the Nijmegen PWA data are also listed. A clear improvement from LO to NLO is observed. 

\begin{table}[h!]
\caption{Scattering length and effective range for the $^1S_0$ and $^3S_1$ partial waves using LO and NLO potentials, compared to the Nijmegen PWA ~\cite{deSwart:1995ui}.}
\label{Tab:1S0-3S1ERE}
\begin{ruledtabular}
\begin{tabular}{ccccc}
   & & LO  & NLO & Nijmegen PWA \\
\hline 
$^1S_0$  & $a$ [fm]  &  $ -21.173$ -- $-20.619$            &  $ -23.179$ -- $-23.287$     & $-23.739$\\
 & $r$ [fm]   &  $1.654$ --  $1.856 $            &  $2.548 $ -- $2.574$      & $2.680$ \\
\hline 
$^3S_1$ &  $a$ [fm]    &  $5.454$ -- $5.473$        &  $5.444$ -- $5.451$     & $5.420$ \\
&  $r$ [fm]    &  $1.446$ -- $1.552$          &  $1.697$ -- $1.708$    & $1.753$ \\
\end{tabular}
\end{ruledtabular}
\end{table}

\subsection{P-waves}
\begin{figure}[t]
  \includegraphics[width=0.8\textwidth]{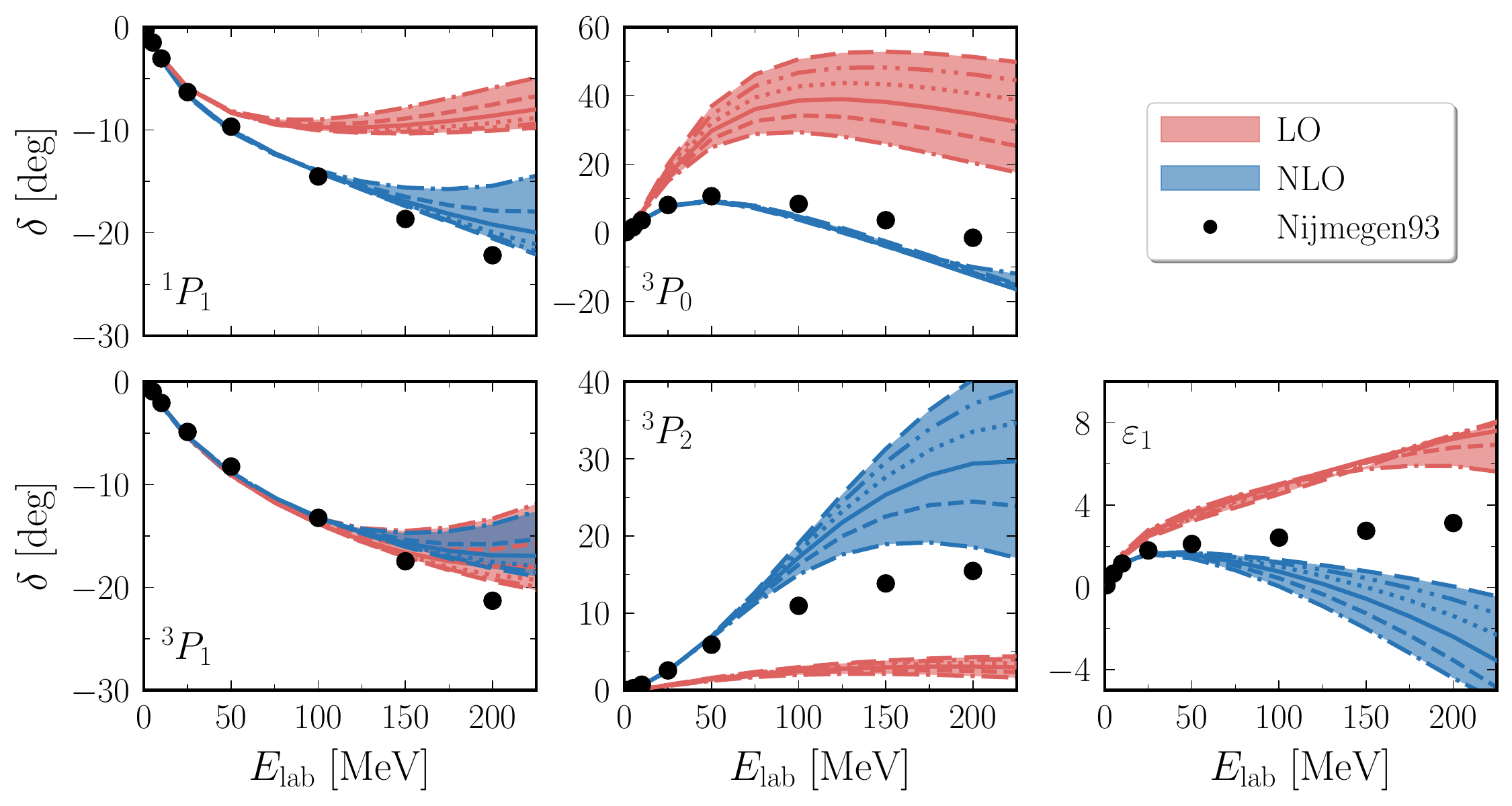}
  \caption{$P$-wave neutron-proton phase shifts and mixing angle $\epsilon_1$ versus the laboratory energy. Notations are given in Fig.~\ref{Fig:Swave}. 
  }
  \label{Fig:Pwave}
\end{figure}

For the P-waves, in Fig.~\ref{Fig:Pwave} we present the LO and NLO phase shifts up to laboratory energy of $200$ MeV. Visible improvements are observed at NLO for the $^1P_1$, $^3P_0$, and $^3P_2$ waves. As for the $^3P_1$ wave, a rather good description has accomplished at LO, thus the very similar result is observed by including the NLO correction.
 
 \begin{figure}[t]
  \includegraphics[width=0.8\textwidth]{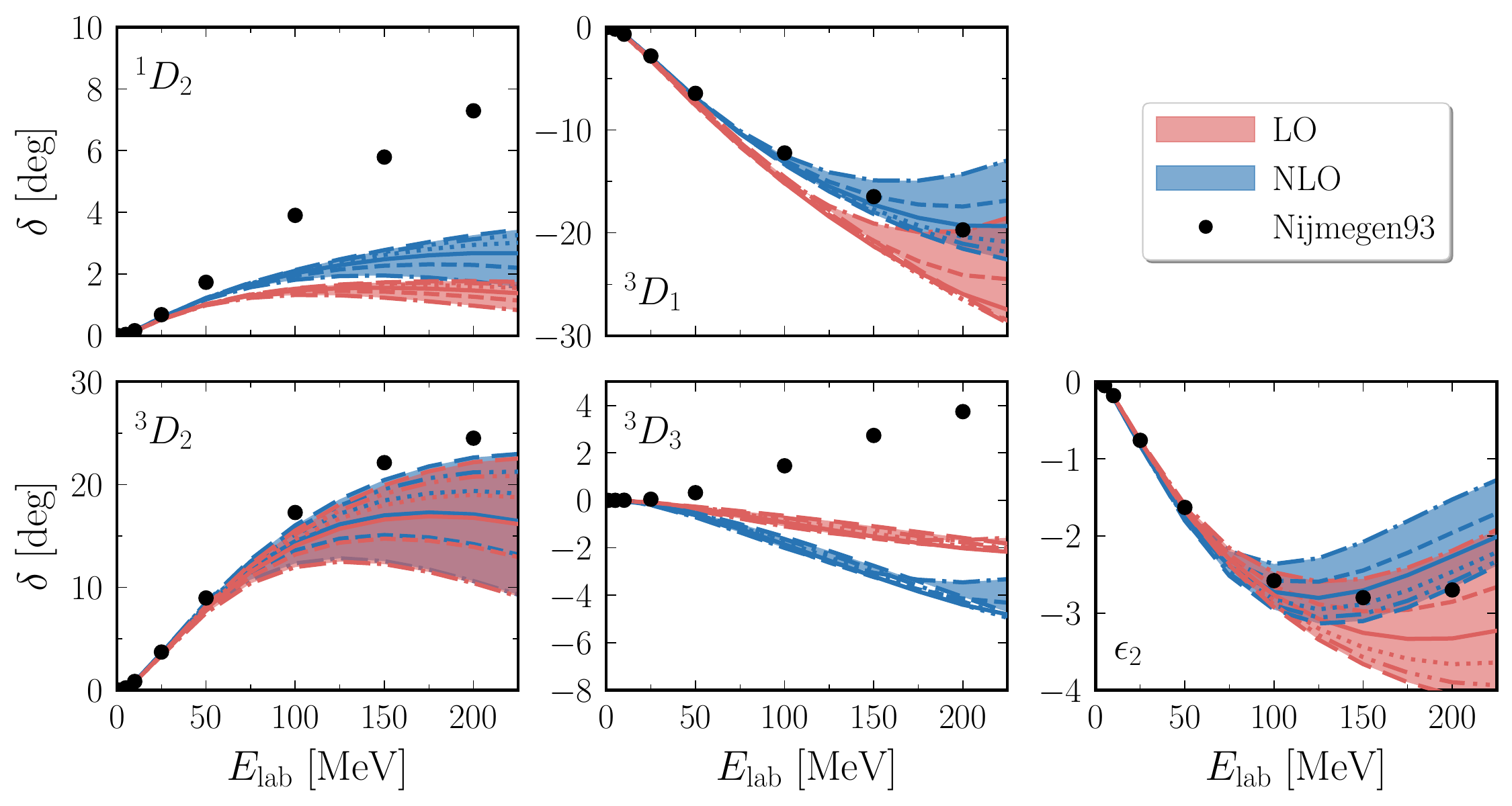}
  \caption{$D$-wave neutron-proton phase shifts and mixing angle $\epsilon_2$  versus the laboratory energy. Notations are given in Fig.~\ref{Fig:Swave}. 
  }
  \label{Fig:Dwave}
\end{figure}

\begin{figure}[h!]
\includegraphics[width=0.8\textwidth]{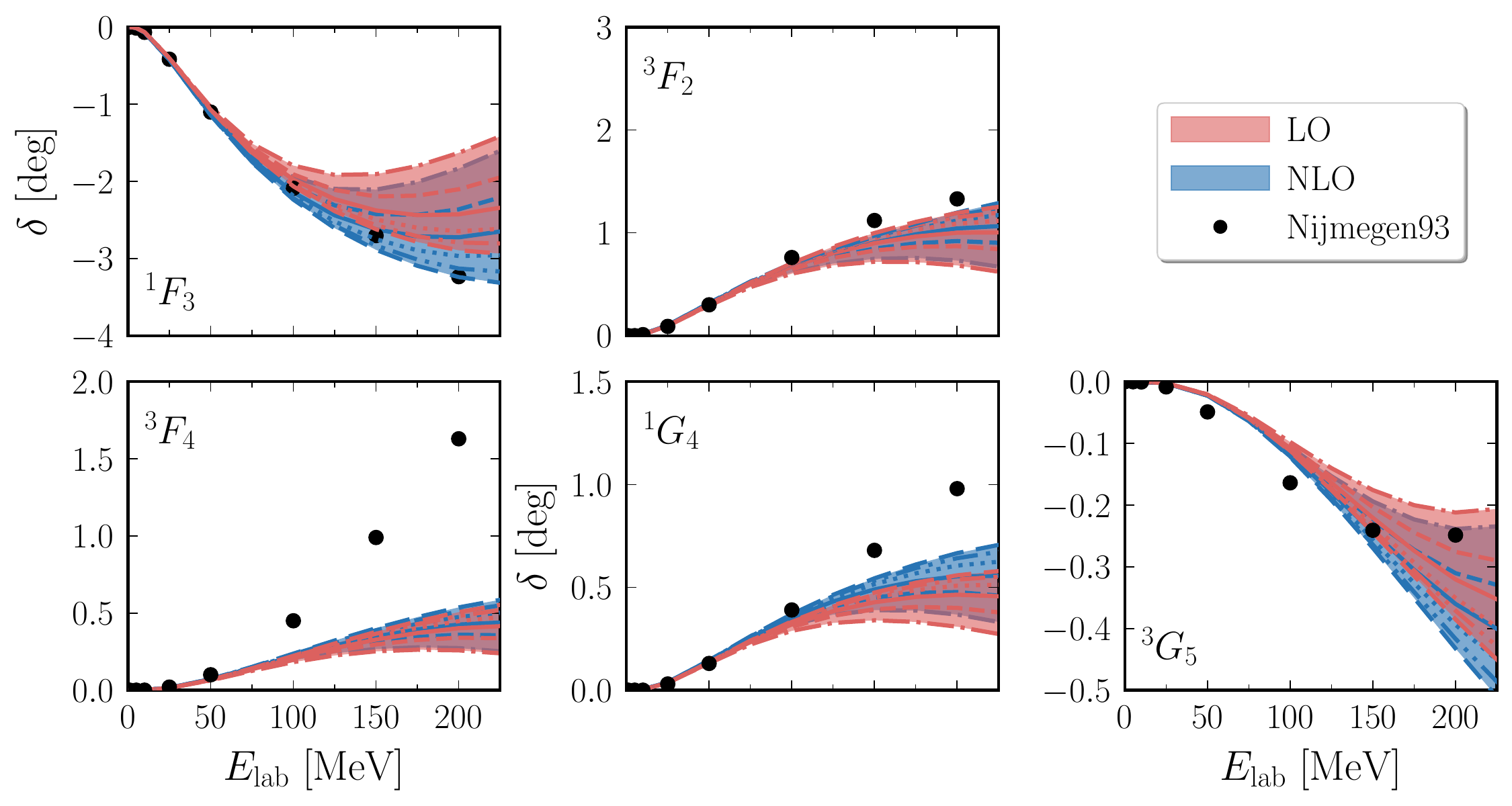}
\caption{Selected $F$- and $G$-wave neutron-proton phase shifts versus the laboratory energy. Notations are given in Fig.~\ref{Fig:Swave}. 
  }
\label{Fig:FDwaves}
\end{figure}

\subsection{D-waves and selected higher partial waves}
In the following we present the results of peripheral phase shifts by solving the scattering equation up to NLO. Since there are no contributions from the contact terms, one can investigate effects of non-perturbative re-summation of the potential on those high partial waves.
First, in Fig.~\ref{Fig:Dwave} we show the prediction of phase shifts for the $D$ partial waves. 
Similarly to the non-relativistic case, the  effects of the non-perturbative re-summation of the leading TPE potential is too weak, which leads to slight improvement of the description of the $^1D_2$ and $^3D_2$ phase shifts, in comparison with the case of OPE potential.
 In comparison with the perturbative calculation in~Ref.~\cite{Ren:2022glg}, the effects of non-perturbative re-summation in both waves is also small at NLO, while the non-perturbative treatment makes the $^3D_3$ potential weaker than the one given in~Ref.~\cite{Ren:2022glg}, which leads to a relatively worse description of Nijmegen PWA. For the mixing angle $\epsilon_2$, both calculations give similar descriptions. 

Furthermore we present the selected peripheral $F$ and $G$ waves in Fig.~\ref{Fig:FDwaves}. One can see that the results are similar to the ones of the perturbative calculation~\cite{Ren:2022glg}, except the $^3G_5$ phase shift, where a relatively better agreement with PWA is found in the current work. 

\subsection{Deuteron properties}
Finally, we come to the two-nucleon bound state problem. 
One can obtain the deuteron wave functions and binding energy by solving the bound state equation, i.e., the homogeneous part of Eq.~\eqref{Eq:kady}. 
Since our TPE potential at NLO has energy dependence, the iterative method is needed to solve the eigenstate problem.  With the fixed parameters at LO and NLO with the cutoff ranging $\Lambda\in [400,\, 650]$ MeV, our predictions of various deuteron properties is tabulated in Table~\ref{Tab:deutron} in comparison to the experimental data. Note that the deuteron binding energy is not used to adjust the LECs. The predicted results for deuteron properties are globally improved from LO to NLO. The deuteron binding energy deviates by only $\sim 2.5\%$ from the experimental value at NLO. 
It is worth noting that no deeply bound states are found in our  NLO calculation.

\begin{table}[h!]
\caption{
Deuteron properties as predicted by NN potentials up to NLO with the cutoff ranging $\Lambda\in [400,\, 650]$ MeV. Here, $E_d$ denotes the binding energy, $Q_d$ the quadrupole moment, $\eta_d$ the asymptotic $D/S$ ratio, $r_d$ the root-mean-square matter radius, $A_S$ the strength of the asymptotic $S$-wave normalization, and $P_D$ the $D$-state probability. }
\label{Tab:deutron}
\begin{ruledtabular}
\begin{tabular}{l cc c }
   & LO  & NLO  & Exp. \\
\hline 
$E_d~[\mathrm{MeV}]$           & $-2.0456$ -- $-1.9602 $ & $-2.1716$ -- $-2.1277$  & $-2.2254575 (9)$ \\
$Q_d~[\mathrm{fm}^2]$          & $0.2993$ -- $0.3102$   &  $0.2731$ -- $0.2738$ &  $0.2859(3)$ \\
$\eta_d$                                  &  $0.0265$ -- $0.0277$  &  $0.0254$ -- $0.0255$ &  $0.0256(4)$ \\
$r_d~[\mathrm{fm}]$               &  $1.9756$ -- $1.9859$  &  $1.9935$ -- $2.0150$ & $1.97507(78)$ \\
$A_S~[\mathrm{fm}^{-1 / 2}]$ &  $0.8024$ -- $0.8289$  &  $0.8718$ -- $0.8790$ &  $0.8846(9)$ \\
$P_D~[\%]$                             &  $4.8270$ -- $7.7444$ &   $2.6003$ -- $3.5318$ & $-$ \\
\end{tabular}
\end{ruledtabular}
\end{table}

\section{Conclusion}\label{sec4}
We studied the low partial waves in the non-perturbative regime by employing the nucleon-nucleon interaction in the framework of TOPT applied to the manifestly Lorentz-invariant ChEFT.  The whole chiral force at NLO is iterated non-perturbatively in the scattering equation by using the Gaussian momentum-space regulator. The unknown contact terms are determined by fitting the corresponding phase shifts with the laboratory energy up to $50$ MeV. We have achieved a reasonable description of the phase shifts in the $S$ and $P$ waves and the deuteron properties at NLO. 

These findings serve as a feasibility study, demonstrating the viability of our chiral NN potential within the TOPT framework for future applications in the few-/many-body calculations. To achieve this goal, we need to remove the energy dependence in our potential as a first step to make it suitable for applications using established {\it ab initio}  methods. 
It is worth to note that in the current study, the relativistic effects and the artifacts of the finite cutoff regulator cannot be quantitatively disentangled. A more systematic analysis of their relative importance will be done in the future.
In the meantime, a natural regularization method of long-range forces, such as the local regularization method in momentum space of Ref.~\cite{Reinert:2017usi}, is worth to be implemented in our scheme. Particularly, the sub-leading TPE potential at NNLO, with large values of $c_{1,...,4}$, is needed to remove the short-range contribution in order to perform the NNLO study in our scheme.
Work along those lines is in progress. 
 
\acknowledgements 
This work has been supported in part by Shandong Provincial Natural Science Fund for Excellent Young Scientists Fund Program (Overseas) with project no. 2025HWYQ-015, by the European Research
Council (ERC AdG Nuclear Theory, Grant No. 885150),
by the MKW NRW under the funding code NW21-024-A, by JST ERATO (Grant No. JPMJER2304) and by JSPS KAKENHI (Grant No. JP20H05636), by the Georgian Shota Rustaveli National Science Foundation (Grant No. FR-23-856), and  by Qilu Youth Scholars Program of Shandong University.
 
\bibliographystyle{apsrev4-2}
\bibliography{ref_non-pert_NN}

\end{document}